\documentclass[12pt]{article}

\usepackage[a4paper,left=30mm,right=20mm,top=20mm,bottom=20mm]{geometry}
\usepackage{graphics}
\usepackage{amsmath}
\usepackage{amssymb}
\usepackage{epsfig}
\usepackage{cite}
\usepackage{colordvi}
\usepackage{color}

\input epsf

\title{On relation between Parton Branching Approach and CCFM evolution}

\author{A.V.~Lipatov$^{1,\,2}$, M.A.~Malyshev$^1$\footnote{e-mail: malyshev@theory.sinp.msu.ru}, H.~Jung$^{3}$}

\begin{document}

\maketitle

\begin{center}

{\it $^1$Skobeltsyn Institute of Nuclear Physics, Lomonosov Moscow State University, 119991 Moscow, Russia}\\
{\it $^2$Joint Institute for Nuclear Research, Dubna 141980, Moscow Region, Russia}\\
{\it $^3$Deutsches Elektronen-Synchrotron, 22603 Hamburg, Germany}

\end{center}

\vspace{0.5cm}

\begin{center}

{\bf Abstract }

\end{center} 

\indent
We consider the associated production of electroweak gauge bosons and charm or beauty quark jets at the LHC using the $k_T$-factorization framework.
We apply the transverse momentum dependent (TMD) parton distributions in a proton 
obtained from the Parton Branching (PB) method as well as from the
Catani-Ciafaloni-Fiorani-Marchesini (CCFM) evolution equation.
For the PB approach, our prescription merges the standard leading order
${\cal O}(\alpha \alpha_s)$ $k_T$-factorization calculations with several tree-level next-to-leading order ${\cal O}(\alpha \alpha_s^2)$ off-shell production amplitudes.
For the CCFM scenario, our consideration is based on the
${\cal O}(\alpha \alpha_s^2)$ off-shell gluon-gluon fusion
subprocess $g^* g^* \to Z/W + Q\bar Q$
and some subleading ${\cal O}(\alpha \alpha_s^2)$ subprocesses involving quark  
interactions, taken into account in conventional (collinear) QCD factorization.
We find that the $W + c$ and $Z + b$ cross sections, calculated within the PB and CCFM-based schemes with the proper choice of leading and next-to-leading subprocesses, 
are in good agreement with each other, thus
establishing a correspondence between these two scenarios.
A comparison with the latest LHC experimental data is
given and the necessity for the 
proper off-shell treatment of the production
amplitudes in determination of the parameters 
of the TMD parton density is demonstrated.

\vspace{1.0cm}

\noindent PACS number(s): 12.38.-t, 12.38.Bx, 13.85.Qk, 14.65.-q, 14.70.Hp

\newpage

\section{Motivation} \indent

A theoretical description of a number of processes at high energies 
and large momentum transfer containing multiple hard scales requires 
so-called transverse momentum dependent (TMD, or unintegrated) 
parton (quark and gluon) density functions\cite{1}.
These quantities encode the non-perturbative information 
on proton structure, including transverse momentum and polarization 
degrees of freedom and are related to the physical cross sections
via different TMD factorization scenarios. The latter
provide the necessary framework to separate hard partonic physics, 
described with a perturbative QCD expansion, from soft hadronic physics.

In the limit of a fixed hard scale and high energy 
the $k_T$-factorization\cite{2} (or high-energy factorization\cite{3}) 
approach is expected to be valid. This approach is mainly
based on the Balitsky-Fadin-Kuraev-Lipatov (BFKL)\cite{4} or 
Ciafaloni-Catani-Fiorani-Marchesini (CCFM)\cite{5} 
evolution equations for the TMD gluon densities in a proton.
The BFKL equation resums large logarithmic terms 
proportional to $\alpha_s^n \ln^n s \sim \alpha_s^n \ln^n 1/x$, important at high energies $s$ (or, equivalently, at small $x$).
The CCFM equation takes into account additional terms 
proportional to $\alpha_s^n \ln^n 1/(1 - x)$ and 
is valid at both low and large $x$.
There are also scenarios to evaluate the TMD parton densities based on the 
conventional Dokshitzer-Gribov-Lipatov-Altarelli-Parisi (DGLAP)\cite{6}
evolution equations, namely the
Kimber-Martin-Ryskin (KMR) prescription\cite{7}
and recently proposed Parton Branching (PB) approach\cite{8,9}.
The KMR approach, currently explored\cite{10} at  
next-to-leading order (NLO), is a formalism to construct the 
TMD parton densities from well-known conventional (collinear) ones
under the key assumption that the
transverse momentum dependence of the parton distributions enters only at the  
last evolution step. 
The PB approach provides an iterative
solution of the DGLAP evolution equations for collinear and TMD parton 
density functions by making use of the concept of
resolvable and non-resolvable branchings and by applying Sudakov form 
factors to describe the parton evolution from one scale to 
another without resolvable branching. The 
splitting kinematics at each branching vertex
is described by the 
DGLAP equations and angular ordering conditions for parton emissions 
can be applied here instead of usual DGLAP ordering in virtuality.
One of the main advantages of the PB approach is that the TMD parton 
densities (and all corresponding non-perturbative parameters)
can be fitted to experimental data, so that the theoretical
predictions, where the parton shower effects are 
already taken into account, can be obtained with no further free 
parameters\footnote{In contrast to the usual parton shower event generators.}.

A number of phenomenological applications of the CCFM evolution equation
is known in the literature (see, for example,\cite{12,13,14,15,16,17,18} and references therein).
Several applications of the PB approach were discussed\cite{19,20}
and comparison between the PB and KMR predictions has been recently 
made\cite{21,21a}.
However, the correspondence between the CCFM and PB scenarios
has been not investigated yet.
One of the main goals of this paper is to compare predictions 
for some QCD processes with the CCFM and PB parton distributions, 
to find a correspondence between these approaches and to 
define conditions, at which such correspondence takes place.
As the reference processes for the study, we consider here
the associated production of electroweak gauge bosons ($W$ and $Z$)
and charm or beauty quark jets at the LHC conditions.
These are the so-called ``rare'' processes which could 
have never been systematically studied at previous accelerators.
We already succesfully applied\cite{17} the CCFM-evolved 
gluon densities to describe first LHC data\cite{22,23} of the
associated $Z + b$ production at 
$\sqrt s = 7$~TeV.
Those calculations were based on the ${\cal O}(\alpha \alpha_s^2)$ 
off-shell gluon-gluon fusion subprocess $g^* g^* \to Z + Q\bar Q$
(where the $Z$ boson further decays into a lepton pair)
and several subleading ${\cal O}(\alpha \alpha_s^2)$
and ${\cal O}(\alpha \alpha_s^3)$ subprocesses involving 
quark-antiquark and quark-gluon interactions,
taken into account within the conventional (collinear) QCD factorization.
Such a scheme allows us to describe LHC experimental data in the
whole transverse momentum range. Here we 
extend the consideration to associated $W + c$ production,
measured\cite{25} by the CMS Collaboration for the first time as a function 
of $W$ decay lepton and/or $c$-jet rapidities  at $\sqrt s = 13$~TeV. 
Thus, in this sense we continue the line of our previous studies\cite{17}. 
In contrast to the CCFM scenario, 
in the PB calculations (as being the DGLAP-based ones)
one has to include usual leading order (LO) ${\cal O}(\alpha \alpha_s)$
subprocesses properly matched with a number of additional higher-order 
terms. Below we perform such calculations and matching procedure 
following the approach applied recently\cite{21} for $c$-jet production 
at the LHC.
The comparison between the results obtained within the TMD
approaches above could be also a general consistency check for 
the $k_T$-factorization phenomenology.

The outline of the paper is the following. In Section~2 we briefly
discuss the CCFM equation and PB approach and 
describe the basic steps of our calculations. 
In Section~3 we present the results of our calculations and discussion.
Our conclusions are summarised in Section~4.

\section{Theoretical framework} \indent

\subsection{CCFM evolution} \indent

The CCFM gluon evolution equation resums large logarithms
$\alpha_s^n \ln^n 1/(1 - x)$ in addition to
BFKL ones $\alpha_s^n \ln^n 1/x$ and introduces angular 
ordering of initial emissions to correctly treat gluon 
coherence effects. In the limit of asymptotic energies,
it is almost equivalent to BFKL, but also similar to the 
DGLAP evolution for large $x$\cite{5}.
In the leading logarthmic approximation, the CCFM equation 
for TMD gluon density ${\cal A}(x,{\mathbf k}_T^2,\mu^2)$
with respect to the evolution (factorization) scale $\mu^2$
can be written as
\begin{equation}
  \displaystyle {\cal A}(x,{\mathbf k}_T^2,\mu^2) = {\cal A}^{(0)}(x,{\mathbf k}_T^2,\mu_0^2) \Delta_s(\mu,\mu_0) + \atop { 
  \displaystyle + \int\frac{dz}{z}\int\frac{dq^2}{q^2}\Theta(\mu-zq)\Delta_s(\mu,zq) \tilde P_{gg}(z,{\mathbf k}_T^2, q^2) {\cal A}\left(\frac{x}{z},{\mathbf k}^{\prime \, 2}_T,q^2\right) },
\end{equation}

\noindent
where ${\mathbf k}_T^\prime = {\mathbf q}(1 - z) + {\mathbf k}_T$
and $\tilde P_{gg}(z,{\mathbf k}^2_T,q^2)$ is the CCFM
splitting function:
\begin{equation}
  \displaystyle \tilde P_{gg}(z,{\mathbf k}^2_T,q^2) = \bar\alpha_s(q^2(1-z)^2) \left[\frac{1}{1-z}+\frac{z(1-z)}{2}\right] + \atop {
  \displaystyle + \bar\alpha_s({\mathbf k}_T^2)\left[\frac{1}{z}-1+\frac{z(1-z)}{2}\right]\Delta_{ns}(z,{\mathbf k}^2_T,q^2) }.
\end{equation}

\noindent
The Sudakov and non-Sudakov form factors read:
\begin{equation}
 \ln \Delta_s(\mu,\mu_0)= - \int\limits_{\mu_0^2}^{\mu^2}\frac{d\mu^{\prime \, 2}}{\mu^{\prime \, 2}}\int\limits_0^{z_M=1-\mu_0/\mu^\prime}dz\,\frac{\bar\alpha_s(\mu^{\prime \, 2}(1-z)^2)}{1-z},
\end{equation}
\begin{equation}
\ln \Delta_{ns}(z,{\mathbf k}_T^2, {\mathbf q}_T^2) = -\bar\alpha_s({\mathbf k}_T^2)\int\limits_0^1\frac{dz^\prime}{z^\prime}\int\frac{dq^2}{q^2}\Theta({\mathbf k}_T^2-q^2)\Theta(q^2-z^{\prime\,2} {\mathbf q}^2_T).
\end{equation}

\noindent
where $\bar \alpha_s = 3 \alpha_s/\pi$. 
The first term in the CCFM equation, which is the initial 
TMD gluon density multiplied by the Sudakov form factor, 
corresponds to the contribution of non-resolvable branchings between 
the starting scale $\mu_0^2$ and scale $\mu^2$.
The second term describes the details of the QCD evolution 
expressed by the convolution of the CCFM gluon splitting function
with the gluon density and the Sudakov form factor. The 
theta function introduces the angular ordering condition.
The evolution scale $\mu^2$ is defined by the maximum allowed
angle for any gluon emission\cite{5}.
A similar equation also can be written\cite{28} for valence quark 
densities\footnote{The sea quark distributions are not defined in 
CCFM. However, they can be obtained from the gluon ones in 
the last gluon splitting approximation, see\cite{27}.} (with replacement of the gluon splitting function by the 
quark one). 
Usually, the initial TMD gluon and valence quark distributions
are taken as
\begin{equation}
  x {\cal A}_g^{(0)}(x, {\mathbf k}_T^2, \mu_0^2) = Nx^{-B}(1-x)^C\exp(-{\mathbf k}_T^2/\sigma^2),
\end{equation}
\begin{equation}
  x {\cal A}_{q_v}^{(0)}(x, {\mathbf k}_T^2, \mu_0^2) = x q_v(x, \mu_0^2) \exp(-{\mathbf k}_T^2/\sigma^2)/\sigma^2,
\end{equation}

\noindent
where $\sigma=\mu_0/\sqrt{2}$ and $q_{v}(x,\mu^2)$ is the 
standard (collinear) density function.
The parameters of the initial TMD parton 
distributions can be fitted from the collider 
data (see, for example,\cite{28,29}).

The CCFM equation can be solved numerically using 
the \textsc{updfevolv} program\cite{30},
and the TMD gluon and valence quark densities can be 
obtained for any $x$, ${\mathbf k}_T^2$ and $\mu^2$ values.
The main advantage of this approach is the ease of 
including higher-order radiative corrections (namely, a part 
of NLO + NNLO +... terms corresponding to the initial-state real gluon
emissions) even within LO. 
More details can be found, 
for example, in review\cite{1}.

\subsection{Parton Branching approach} \indent

The Parton Branching approach was introduced\cite{8,9} to 
treat the DGLAP evolution. The method 
provides an iterative solution of the evolution equations
and agrees with the usual methods to solve the DGLAP
equations for inclusive distributions at the NLO and NNLO.
It allows one to take into account simultaneously soft-gluon 
emission at $z \to 1$ and transverse momentum $\mathbf {q}_T$
recoils in the parton branchings along the QCD cascade.
The latter leads to a natural determination 
of the TMD quark and gluon densities.
A soft-gluon resolution scale $z_M$ is introduced 
to separate resolvable and non-resolvable emissions,
which are treated via the DGLAP splitting functions $P_{ab}(\alpha_s, z)$
and Sudakov form-factors, respectively.
The PB equations for TMD parton densities read:
\begin{equation}
  \displaystyle x{\cal A}_a(x,{\mathbf k}_T^2,\mu^2) = x{\cal A}_a^{(0)}(x,{\mathbf k}_T^2,\mu_0^2)\Delta_a(z_M,\mu,\mu_0) + \atop{ 
  \displaystyle + \sum_b\int\limits_x^{z_M}dz\int\frac{dq^2}{q^2}\Theta(\mu^2- q^2)\Theta(q^2-\mu_0^2) \Delta_a(z_M,\mu,q)
P_{ab}(\alpha_s, z) \frac{x}{z}{\cal A}_b\left(\frac{x}{z},{\mathbf k}_T^{\prime \,2},q^2\right) },
\end{equation}

\noindent
where ${\mathbf k}_T^\prime = {\mathbf q}(1 - z) + {\mathbf k}_T$.
The Sudakov form factors are defined as
\begin{equation}
  \ln \Delta_a(z_M,\mu,\mu_0)= -\sum_b\int\limits_{\mu_0^2}^{\mu^2}\frac{d\mu^{\prime \, 2}}{\mu^{\prime \,2}}\int\limits_0^{z_M}dz\,z\,P_{ba}(\alpha_s(\mu^{\prime \, 2}),z).
\end{equation}

\noindent
The evolution scale $\mu^2$ can be connected with the 
angle of emitted parton with respect to the beam direction, that
leads to the well-known angular ordering condition, $\mu = |{\mathbf q}_T|/(1 - z)$.
The dependence on the $z_M$ 
falls out when this angular ordering condition is applied and $z_M$ is 
large enough.
The initial TMD parton distributions are taken in a factorized form
as a product of collinear quark and gluon densities and 
intrinsic transverse momentum distributions (treated as gaussian 
ones\cite{8,9}), where all the parameters can be fitted from the collider data.
Unlike the CCFM parton distributions, the PB densities have the strong normalization property:
\begin{equation}
  \int {\cal A}_a(x, {\mathbf k}_T^2,\mu^2) d{\mathbf k}_T^2 = f_a(x,\mu^2).
\end{equation}

\noindent
The PB evolution equations can be solved by an iterative Monte-Carlo
method, that results in a steep drop of the parton 
densities at ${\mathbf k}_T^2 > \mu^2$.
It contrasts the CCFM evolution,
where the transverse momentum is allowed to be larger
than the scale $\mu^2$, corresponding to an effective taking into account
 higher-order contributions\footnote{Very recently, 
a method to incorporate CCFM effects into the PB 
formulation was proposed\cite{31}.}.

\subsection{Associated $W^\pm/Z + Q$ production with TMD factorization} \indent

To calculate total and differential cross sections of associated
gauge bosons and heavy quark jet production within the CCFM-based approach, 
we strictly follow the scheme\cite{17}.
In this scheme, the leading contribution comes from the
${\cal O}(\alpha \alpha_s^2)$ off-shell gluon-gluon fusion
subprocess
\begin{gather}
\label{ggZ}
  g^* + g^* \to Z + b + \bar b,\\  
\label{gg}
  g^* + g^* \to W^{-} + c + \bar s.  
\end{gather}

\noindent
In addition to off-shell gluon-gluon fusion, one can take into 
account several ${\cal O}(\alpha \alpha_s^2)$ subprocesses involving quarks in the initial state:
\begin{gather}
\label{qqsZ}
  q + \bar q \to Z + b +\bar b,\\
\label{qqtZ}
  q + b \to Z + b + q,\\
\label{bg4}
  g + b \to Z + b + g
\end{gather}

\noindent
for $Z + b$ production and
\begin{gather}
\label{qqs}
  q + \bar q \to W^- + c + \bar s,\\
\label{qqt}
  q + s \to W^- + c + q,\\
\label{sg4}
  g + s \to W^- + c + g
\end{gather}

\noindent
for $W^- + c$ production. 
Subprocesses for $W^+ + \bar c$ production can be obtained via 
charge conjugation\footnote{The event selection in \cite{25} is organized 
in a way to exclude subprocesses with gluon splitting $g\to c\bar c$, 
so such subprocesses are left out of the consideration.}. 
The quark-induced diagrams may become important at very large 
transverse momenta (or, respectively, at large $x$, which is 
needed to produce large $p_T$ events) where the quarks are 
less suppressed or can even dominate over the gluon density.
Following\cite{17}, the contributions from subprocesses (\ref{qqsZ}), (\ref{qqtZ}), (\ref{qqs}) and (\ref{qqt}),  are taken into account 
using the collinear
DGLAP-based factorization scheme, which provides better 
theoretical grounds in the large $x$ region\footnote{Subprocesses (\ref{bg4}) and (\ref{sg4}) are partly taken into account with the gluon fusion subprocesses (\ref{ggZ}) and (\ref{gg}), respectively, in the $k_T$-factorization approach.}. 
So, we consider a combination of two techniques
with each of them being used at the kinematic conditions
where it is best suitable.
We note that the contributions from the off-shell ${\cal O}(\alpha \alpha_s)$ 
subprocesses, namely,
\begin{gather}
\label{bg}
  b^* + g^* \to Z + b, \\  
\label{sg}
  s^* + g^* \to W^{-} + c  
\end{gather}

\noindent
in the CCFM scheme are covered by gluon-fusion 
subprocesses (10) or (11) and therefore not taken into account to avoid the double counting.
In contrast, in the DGLAP-based PB approach one has to 
take into account the off-shell ${\cal O}(\alpha \alpha_s)$ 
subprocesses (18) or (19) and properly match them
with the higher-order contributions (12) --- (14) or (15) --- (17), 
respectively.
The details of the matching procedure are discussed below (see Sec.~\ref{matching}).

According to $k_T$-factorization prescription, to calculate the 
cross sections of processes under consideration 
we have to convolute the relevant 
partonic cross sections (related with the
off-shell production amplitudes) and TMD
parton densities in a proton:
\begin{equation}
  \sigma = \sum_{a,b}\int dx_1 dx_2 d{\mathbf k}_{1T}^2 d{\mathbf k}_{2T}^2 d\hat\sigma_{ab}^*(x_1,x_2,\mathbf k_{1T}^2,\mathbf k_{2T}^2,\mu^2)\mathcal A_a(x_1,\mathbf k_{1T}^2,\mu^2)\mathcal A_b(x_2,\mathbf k_{2T}^2,\mu^2),
\end{equation}
where $x_1$ and $x_2$ are the longitudinal momentum fractions of 
the initial off-shell partons and $\mathbf k_{1T}^2$ and $\mathbf k_{2T}^2$ 
are their transverse momenta.
The gauge-invariant off-shell production amplitudes for gluon-gluon fusion 
subprocesses (10) and (11) were calculated earlier\cite{32,33} and 
implemented into the Monte-Carlo event generator \textsc{cascade}\cite{34}
and newly developed parton-level 
Monte-Carlo event generator \textsc{pegasus}\cite{35}.
The off-shell amplitudes for quark induced subprocesses
(12) --- (14) and (15) --- (17) can be derived in the 
framework of the reggeized parton approach\cite{36}. 
One can also use Britto-Cachazo-Feng-Witten (BCFW) recursion for off-shell 
gluons\cite{37} and method of auxilliary quarks for off-shell 
quarks\cite{38}, implemented in the 
Monte-Carlo generator \textsc{katie}\cite{39}.
In this study, to calculate the contributions from 
(12) --- (14) and (15) --- (17) 
subprocesses in the PB scheme 
we used the \textsc{katie} tool.

In the present paper we compare the CCFM and PB results 
obtained with JH'2013 set~1\cite{28} and 
PB-NLO-HERAI+II-2018 set~2\cite{20} TMD parton densities in a 
proton\footnote{A comprehensive collection of the available TMD
parton densities can be found in the \textsc{tmdlib} package\cite{40}.}.
The main motivation for our choice is that
the input parameters of both these distributions were obtained in exactly the same way: from
the best description of precision 
DIS data on the proton 
structure functions $F_2$ with exactly the same 
angular ordering conditions (see\cite{20,28} for more information).
For the conventional quark and gluon densities we
use the MMHT'2014 (LO) set\cite{41}. 

\section{Numerical results} \indent

Before we show the results of our calculations, we list the input
parameters. 
We use two-loop running strong coupling formula with $n_f=5$ massless 
quark flavours and take $\Lambda_\text{QCD}=200$~MeV in CCFM case 
and $\Lambda_\text{QCD}=118$~MeV for PB distributions~\cite{20,28}. 
The QED running coupling is applied with $\alpha(m_Z^2) = 1/128$.
The electroweak bosons masses were taken as $m_W = 80.385$~GeV 
and $m_Z = 91.188$~GeV\cite{42}. 
As it is often done, we keep the factorization and renormalization scales to be 
equal to the gauge boson mass. However, in the CCFM scheme
we use a different value for factorization 
scale $\mu_F^2= \hat s + {\mathbf Q}_T^2$, where $\hat s$ and 
$\mathbf Q_T$ are the energy of the scattering subprocess and 
transverse momentum of the incoming off-shell gluon pair. 
The definition of $\mu_F$ is unusual and dictated by the CCFM 
evolution algorithm\cite{28}. 

\pagebreak
\subsection{Matching ${\cal O}(\alpha \alpha_s)$ and ${\cal O}(\alpha \alpha_s^2)$ terms in the PB approach} \indent
\label{matching}

As it was already mentioned above, we supplemented the LO
contributions~(18) or (19) with off-shell partons in the PB calculations
by the tree-level $\mathcal O(\alpha_S^2)$ corrections
(13) and (14) or (16) and (17)
from the emission of one additional parton.
However, as it is well known, a problem of possible double counting 
can occur when mixing different final states.
Let us consider the $Z + b$ production (of course, the same 
arguments apply for $W + c$ case).
Here, the off-shell subprocess (18) partially includes 
contributions from (13) and (14) due to initial state parton radiation,
that can result in substantial double counting
if these contributions are summed up.
To avoid this double counting we limit the 
integration over the transverse momenta of the incoming off-shell quark and gluon 
in the factorization formula (20) for the LO 
subprocess (18) from 
above with some value $k_T^\text{cut}$, so 
$|{\mathbf k}_{T1}| < k_T^\text{cut}$ and 
$|{\mathbf k}_{T2}| < k_T^\text{cut}$.
Thus, one removes jets, originating from the 
initial state radiation and being harder than the initial partons. 
The latter, however, could be covered by the subprocesses~(13) and~(14), 
if we choose there only the events with final gluons and light quarks, 
having transverse momenta $p_T$ larger, that the cut scale $k_T^\text{cut}$.
In this way, therefore, we can almost avoid the double counting region.

Of course, the value of $k_T^{\rm cut}$ is not universal but is
depending on the process. 
In order to determine $k_T^{\rm cut}$
we have calculated the differential cross sections of the LO
subprocesses~(18) or (19) as a function of initial gluon
transverse momentum and  
differential cross sections of the $\mathcal O(\alpha_S^2)$ subprocesses~(13) or (14)
as a function of the produced gluon transverse momentum 
$p_T$, see Fig.~1.
These calculations were performed in the fiducial kinematical region
covered by the ATLAS\cite{23} and CMS\cite{25} experiments (see below).
So, fixing the $k_T^{\rm cut}$ at some value would mean 
that we keep the contribution from the $\mathcal O(\alpha_S)$ subprocesses 
lying to the left from the vertical line with $k_T=k_T^\text{cut}$ and 
complement it with the contribution from the $\mathcal O(\alpha_S^2)$ subprocesses 
right to the vertical line. 
The resulting matched $p_T$ distribution will have a step-like 
discontinuous behaviour at $k_T=k_T^\text{cut}$. 
A reasonable choice for the $k_T^\text{cut}$ would be then the one, 
with the step being small. We find, that this can be achieved with
$k_T^\text{cut} \simeq 30$~GeV for associated $Z + b$-jet production
and $k_T^\text{cut} \simeq 15$~GeV for $W + c$-jet 
production.
As one can see from Fig.~1, this choice will lead to the 
continuous merged transverse momentum
distributions.

To investigate the dependence of PB predictions
on the $k_T^\text{cut}$ value in more details we calculated the ratios of 
fiducial cross sections 
$\sigma_{\rm PB}(Z + b)/\sigma_{\rm CCFM}(Z + b)$
and $\sigma_{\rm PB}(W + c)/\sigma_{\rm CCFM}(W + c)$
as a function of $k_T^\text{cut}$ (note that the denominators in these 
ratios do not depend on the $k_T^\text{cut}$). 
Our results are shown in Fig.~2.
We find that the matched PB predictions are rather 
stable with variation in $k_T^\text{cut}$: both the $Z + b$-jet and
$W + c$-jet cross sections 
change less than $5$\% if $k_T^\text{cut} \geq 10$~GeV or 
$k_T^\text{cut} \geq 20$~GeV, respectively.
The dependence on $k_T^\text{cut}$
is smaller than the scale uncertainties of our 
calculations (see estimation below), 
so we employ the matching value $k_T^\text{cut} = 30$~GeV 
for $Z + b$ calculations and $k_T^\text{cut} = 15$~GeV  
for $W + c$ production in our numerical calculations.

One can see that with the appropriate choice of 
$k_T^\text{cut}$, as discussed above, the fiducial cross sections 
calculated in the PB approach are very close to the ones, 
obtained in the CCFM scheme.
The correspondence between these approaches
is investigated in detail in the next Section.

\subsection{Comparison with the LHC data} \indent

We are now in a position to present the results of
our simulations and to confront them
with the latest LHC data.

The measurements of the associated production of 
$Z$ bosons and beauty jets have been carried out by 
the ATLAS\cite{22} and CMS\cite{23} Collaborations and refer 
to the following categories: $Z$ bosons produced in 
association with one beauty jet, $Z$ bosons produced in association 
with two beauty jets, $Z$ bosons associated with any number 
of $b$-jets and $Z$ bosons produced in association with 
explicitly reconstructed $b$-hadrons.
The data on the associated production of $W$ bosons and 
one charmed jet were reported by the CMS Collaboration
very recently for the first time\cite{25}.
In the present study we concentrate on the 
production of gauge bosons associated with one 
heavy quark jet.


The ATLAS Collaboration has collected the data on $Z + b$-jet 
production at the center-of-mass energy$\sqrt s = 7$~TeV\cite{22}. 
Both leptons
originating from the $Z$ boson decay are required to 
have $p_T^l > 20$~GeV and $|\eta^l| < 2.4$, the
lepton pair invariant mass lies in the 
interval $76 < M^{ll} < 106$~GeV, the beauty jets 
are required to have $p_T^b > 20$~GeV and $|\eta^b| < 2.4$.
The measurement of $W + c$-jet production at LHC was made by the CMS 
Collaboration\cite{25} at $\sqrt{s}=13$~TeV
and the fiducial region was defined with the following cuts:
the transverse momentum of the $c$-quark $p_T^c>5$~GeV,  
transverse momentum and pseudo-rapidity of the muon 
originating from $W$ decay should be $p_T^\mu>26$~GeV and 
$\eta^\mu<2.4$. The transverse mass of the $W$ boson
should be $m_T>50$~GeV. 

We start from $Z + b$ production.
In Fig.~3 we present the predictions for the $Z$ boson
rapidity and transverse momentum distributions
in comparison with the ATLAS\cite{22} data.
The solid and dashed histograms corresponds to the results
obtained with the JH'2013 set~1 
and PB-NLO-HERAI+II-2018 set~2 parton densities.
The shaded band represents the scale uncertainties
of our PB-based calculations, which have been estimated by varying the 
scales $\mu_R$ and $\mu_F$ by a factor
of $2$ around their default values
\footnote{The uncertainties, connected with the determination of PB TMD parameters are typically much less, than the estimated scale uncertainties and are not taken into account in this work.}.
For comparison, we also show the conventional (collinear) NLO pQCD 
predictions, taken from\cite{22}, calculated with \textsc{mcfm} generator~\cite{42a}. 
One can see that the $Z$ boson rapidity distribution show almost perfect agreement between 
the CCFM and PB approaches, demonstrating the consistency between the CCFM and PB approaches. 
The cross sections are lower, 
than the ones, obtained in the collinear approach. However, the $k_T$-factorization based calculations are in better 
agreement with the data, though slightly overestimating the ATLAS data in 
the central region. This overestimation is, however, covered by the uncertainties of our calculations.
More information can be obtained with the $p_T^Z$-distributions. 
The CCFM and PB-based calculations give almost the same very
good description of the ATLAS data at small $p_T^Z$, 
while at $p_T^Z\gtrsim 100$~GeV the PB-based cross section 
lies significantly higher, than the CCFM one, and is in better agreement with the data. 
The reason for that is a more accurate treatment of quark-initiated subprocesses in the PB approach, including also contributions from subprocesses (\ref{bg4}) and (\ref{sg4}).
We would like to note, that both approaches describe the
ATLAS data generally better, than the NLO collinear factorization 
results, especially at low $p_T^Z$. The scale uncertainties, estimated for PB scheme, are even less, than the ones of collinear NLO predictions. 

Now we turn to the $W + c$-jet production. Our results are
shown in Fig.~4, where we plot the decay muon rapidity 
distributions for both $W^++\bar c$ and $W^-+c$ events
measured by the CMS Collaboration\cite{25}.
As in the $Z+b$ production, one can see, that the results, obtained with the CCFM and PB approaches agree very well with each other. This confirms once again the consistency between these approaches. However, our predictions are lower, than the collinear predictions, which is in 
contrast to the $Z + b$-jet production.
To explain the reason of the observed underestimation,
let us consider the relevant differential cross sections  
as functions of longitudinal momentum fractions and 
transverse momenta of incoming partons. We show these cross-sections on Figs.~5 and~6.
As one can see, the $W + c$-jet production is dominated 
by smaller $x$ and broader $k_T$ regions in comparison 
with the $Z + b$-jet production case.
The corresponding off-shell production amplitudes are 
known to be supressed in the large $k_T$ domain~\cite{43}. 
We demonstrate this effect in Fig.~7 (left),
where the ratios of the reduced 
cross sections of (18) and (19) calculated with 
off-shell and on-shell production amplitudes are presented.
We find that at $k_T\sim 20$~GeV the off-shell 
amplitude becomes greatly suppressed.
However, a large part of the $W + c$ events 
comes from the region of $k_T\sim 20$~GeV, as one can see from Fig.~6.
Since the considered TMD parton densities were fitted 
with on-shell matrix elements (see\cite{20,28}), the suppression 
results in the drop of the total cross section, 
thus leading to the observed underestimation. The observed flat behavior at relatively low $k_T$ is connected with the kinematical cuts applied in the CMS analysis~\cite{25}. To investigate it in more details we repeat the calculations with different cuts on the transverse momentum of the produced $c$-quark without any other restrictions on the phase space. One can see, that with increasing the $p_T^\text{cut}$ the plateau continuates until a larger value of $k_T\sim p_T^\text{cut}$. In the case of a small value of $p_T^\text{cut}$ we obtain a steep behaviour starting from practically zero $k_T$. A similar observation was made in~\cite{43} for charm and beauty quark photoproduction at HERA, where the heavy quark mass played the role of $p_T^\text{cut}$.
Thus, we conclude that to describe the overall  
normalization of 
$W + c$-jet production\cite{25}
the appropriate fit of the parameters of considered 
TMD parton distributions 
with proper off-shell treatment of production amplitudes
is needed.

Finally, we also make a prediction for the $p_T^W$-distribution in the $W+c$ production case (Fig.~\ref{Fig5}). One can see, that the CCFM and PB approaches result in different shapes of the distribution, however, the position of the peak remains the same. Like in the case of $Z+b$ production, the difference in the shapes can be explained by a more accurate treatment of the quark-initiated subprocesses within the PB calculations.


\section{Conclusion} \noindent

We have studied the associated production of $Z$ and $W$ 
bosons and charm or beauty quark jets at the LHC conditions
using the TMD factorization framework.
We have applied the TMD parton distributions in a proton 
obtained from the recent PB method as well as from the
CCFM evolution equation.
For the PB approach, our prescription merges the
${\cal O}(\alpha \alpha_s)$ calculations 
with several tree-level 
${\cal O}(\alpha \alpha_s^2)$ off-shell production amplitudes.
For the CCFM scenario, our consideration is based on the
${\cal O}(\alpha \alpha_s^2)$ off-shell gluon-gluon fusion
subprocess $g^* g^* \to Z^0/W^\pm Q\bar Q$
and some subleading ${\cal O}(\alpha \alpha_s^2)$ 
subprocesses involving 
quark interactions, taken into account in conventional 
QCD factorization.
We have found that the $W + c$ and $Z + b$ cross sections, 
calculated within the PB and CCFM-based schemes with the proper choice of leading and next-to-leading subprocesses in the $k_T$-factorization, 
are in good agreement with each other. Thus
we have established a correspondence between these two scenarios.
We have demonstrated the necessity for the 
proper off-shell treatment of the production
amplitudes in determination of the parameters 
of the TMD parton dentities in a proton.


\begin{figure}
\begin{center}
\includegraphics[width=7.0cm]{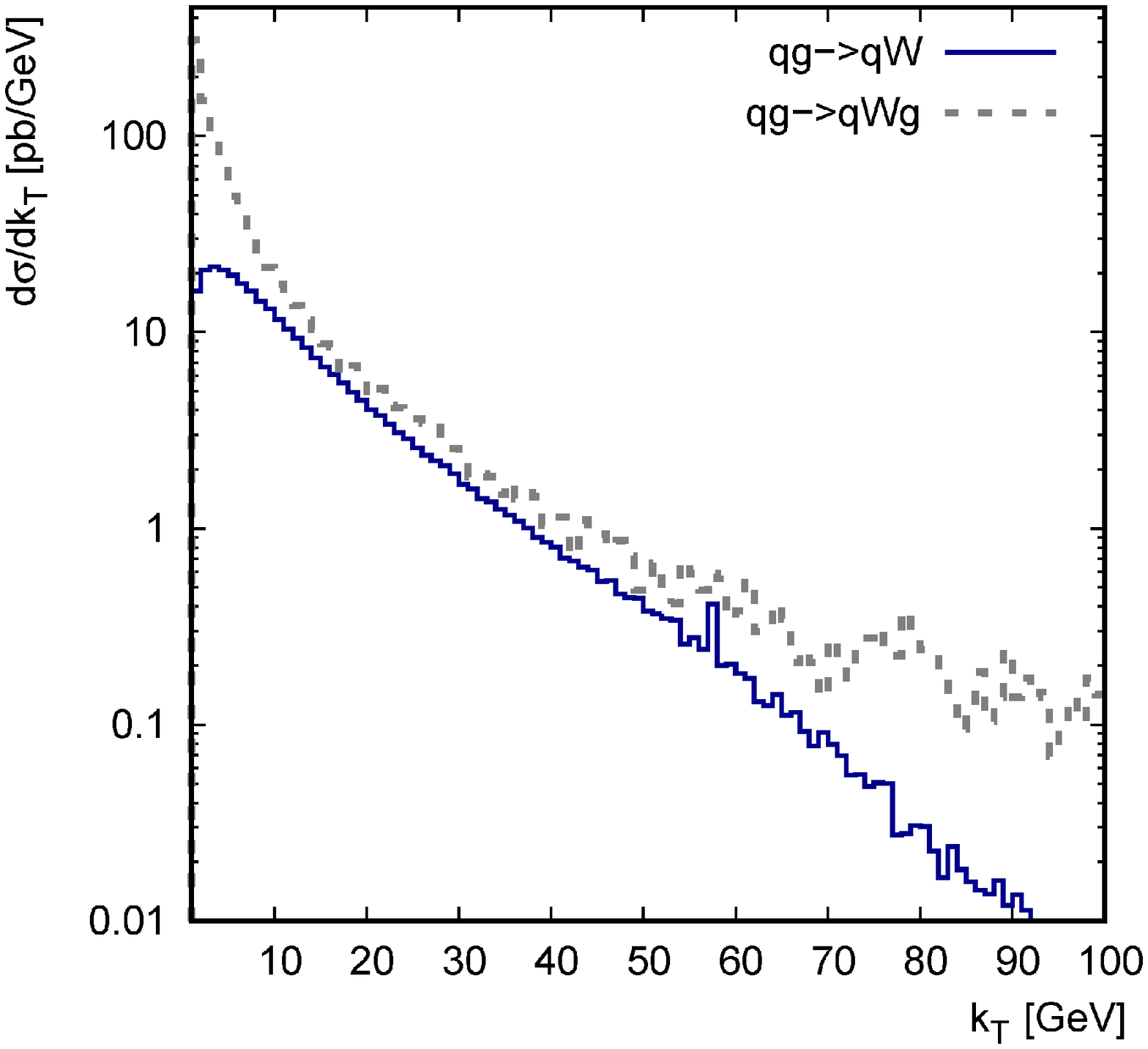}
\includegraphics[width=6.7cm]{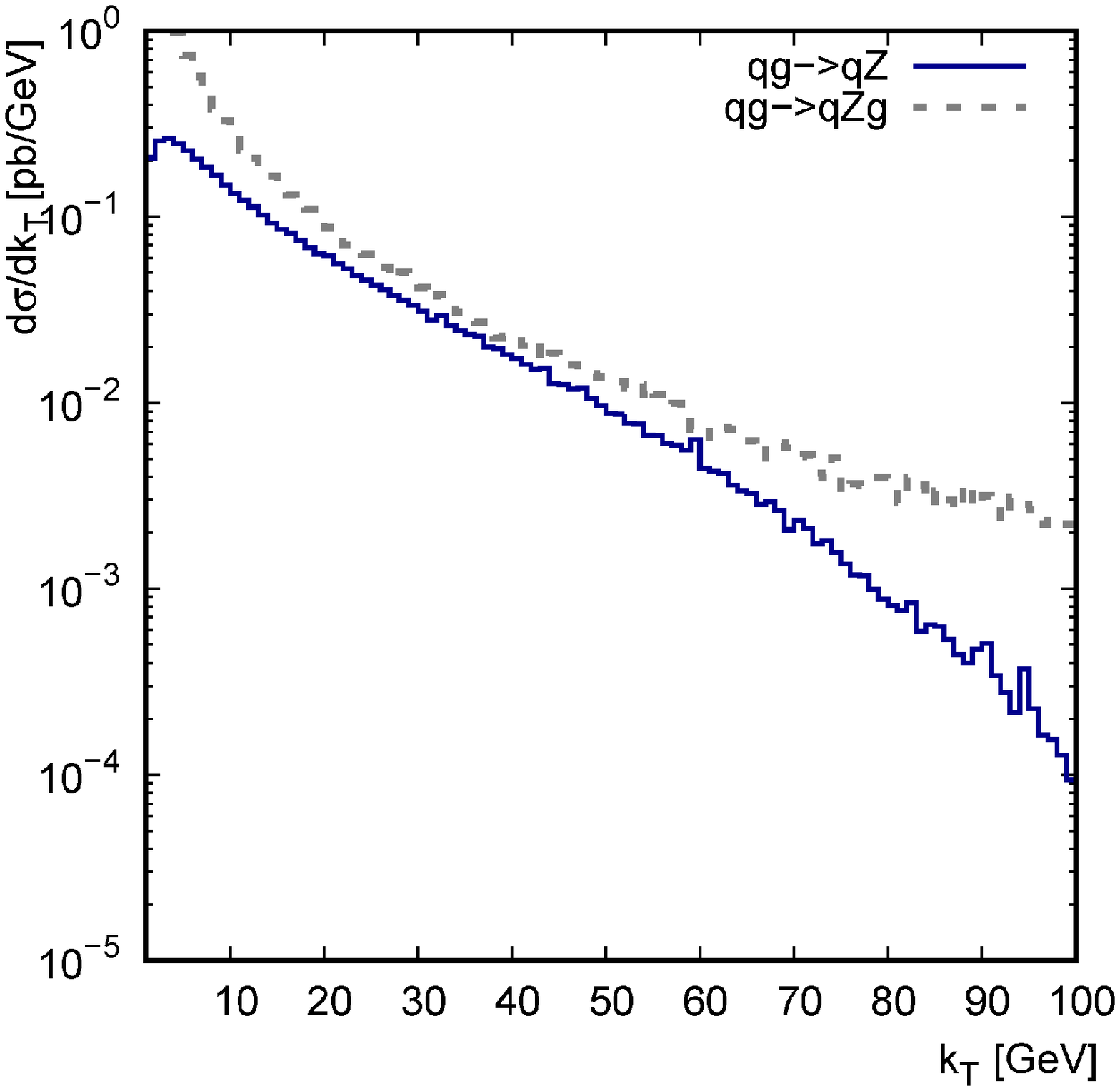}
\caption{Left panel: distribution for $W^-+c$ production in channel~(\ref{sg}) as a function of the transverse momentum of one of the initial partons $k_T$ (solid line) and the same distribution for the subprocess~(\ref{sg4}) as a function of the transverse momentum of the final gluon $p_T$ (dashed line). Right panel: the same distributions for $Z+b$ production in channels~(\ref{bg}) (solid line) and~(\ref{bg4}) (dashed line).}
\label{Fig1}
\end{center}
\end{figure}

\begin{figure}
\begin{center}
\includegraphics[width=7.0cm]{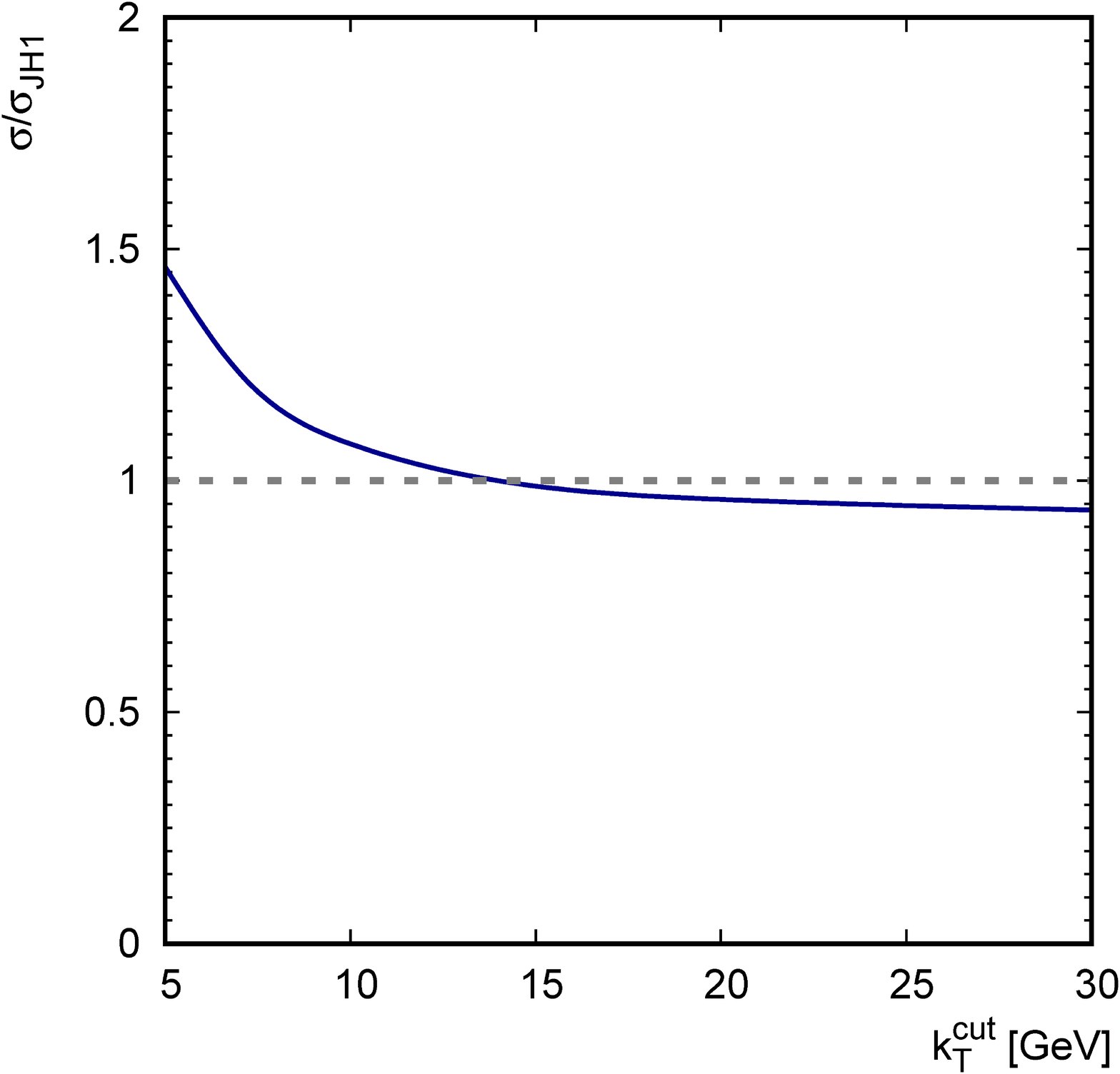}
\includegraphics[width=7.0cm]{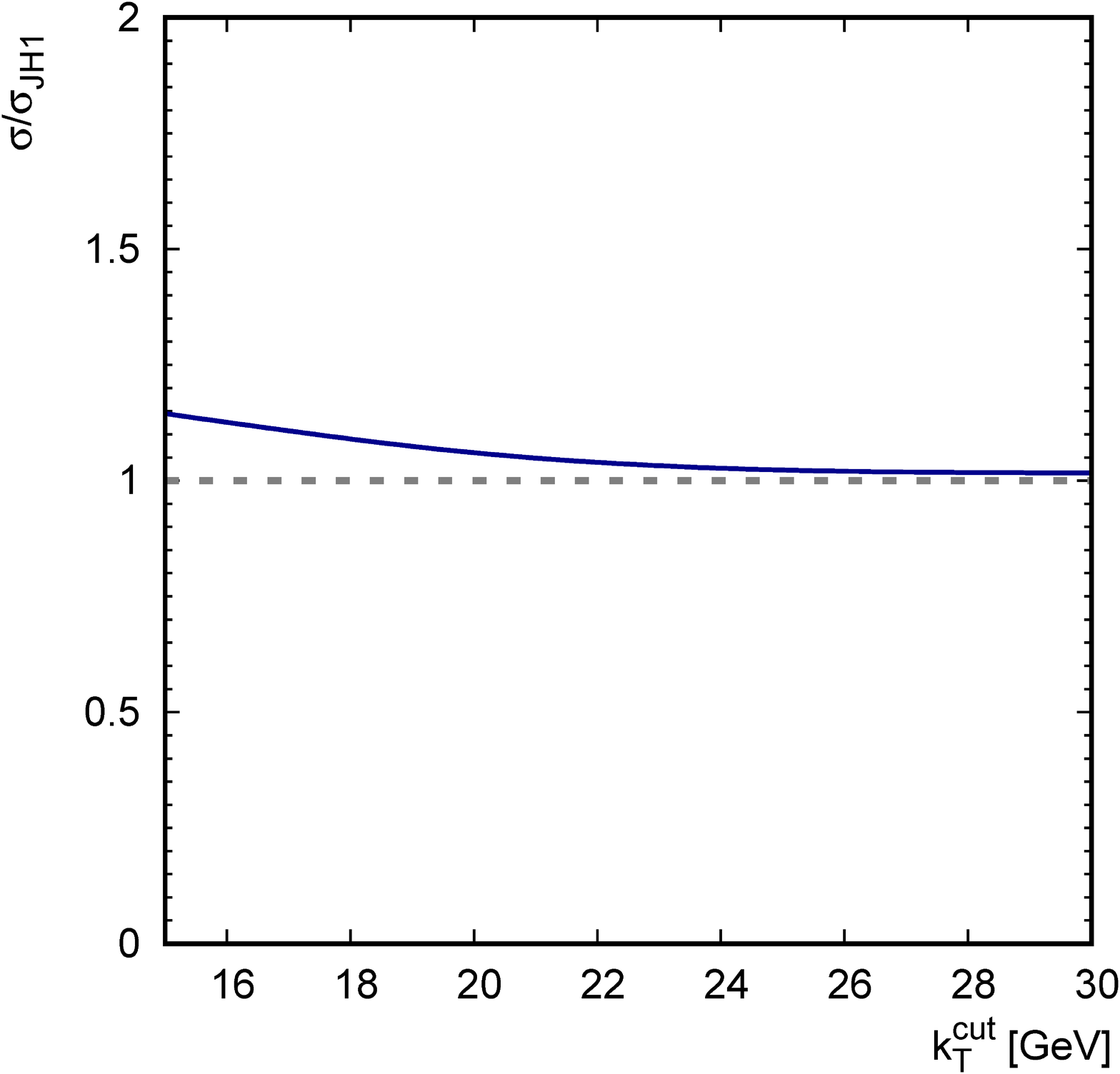}
\caption{Dependence of the fiducial cross section normalized to the cross section, obtained with JH'2013 set 1 TMD parton densities, as a function of $k_T^\text{cut}$. Left panel: $W^-+c$ production. Right panel: $Z+b$ production}
\label{Fig2}
\end{center}
\end{figure}


\begin{figure}
\begin{center}
\includegraphics[width=7.0cm,height=6.9cm]{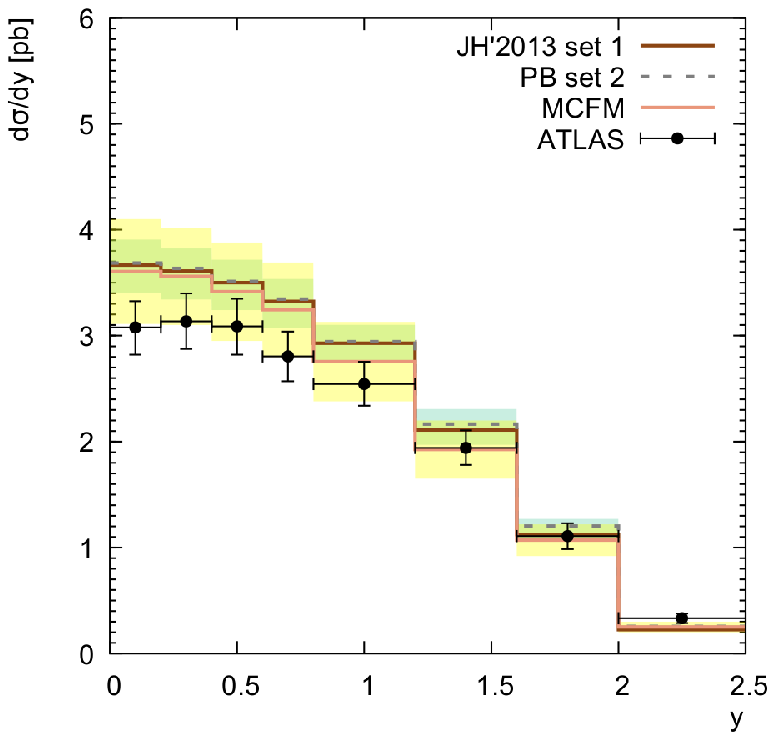}
\includegraphics[width=7.0cm,height=6.9cm]{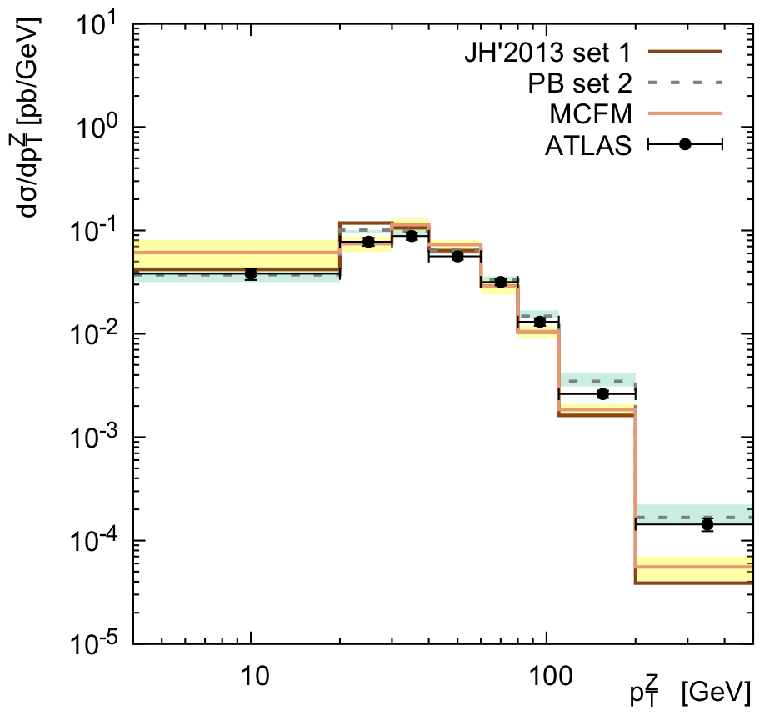}
\caption{Cross sections of $Z+b$-production as functions of the $Z$-boson rapidity (left) and transverse momentum (right). The solid brown line corresponds to the the CCFM approach with JH'2013 set~1 TMD parton density; the grey dashed line corresponds the PB approach with PB-NLO-HERAI+II-2018 set~2 TMD parton densities; the solid pink line corresponds to the collinear factorization approach in NLO. The data are from ATLAS~\cite{22}.}
\label{Fig4}
\end{center}
\end{figure}

\begin{figure}
\begin{center}
\includegraphics[width=7.0cm]{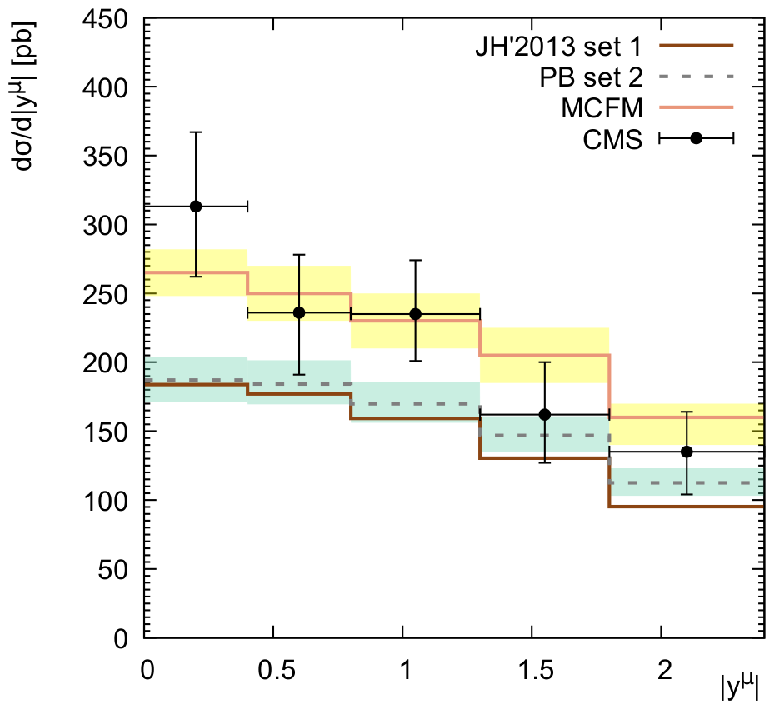}
\includegraphics[width=7.0cm]{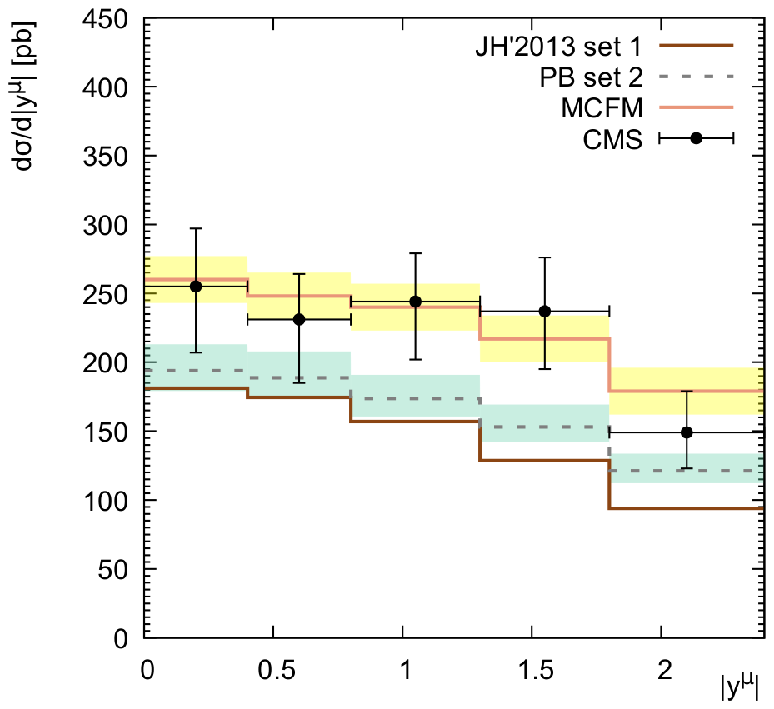}
\caption{Cross sections of $W+c$ production as functions of the decay muon rapidity. The notations are the same, as on the previous figure. Left panel: $W^++\bar c$ production case. Right panel: $W^-+c$ production case. The data are from CMS~\cite{25}.}
\label{Fig3}
\end{center}
\end{figure}


\begin{figure}
\begin{center}
\includegraphics[width=7.0cm]{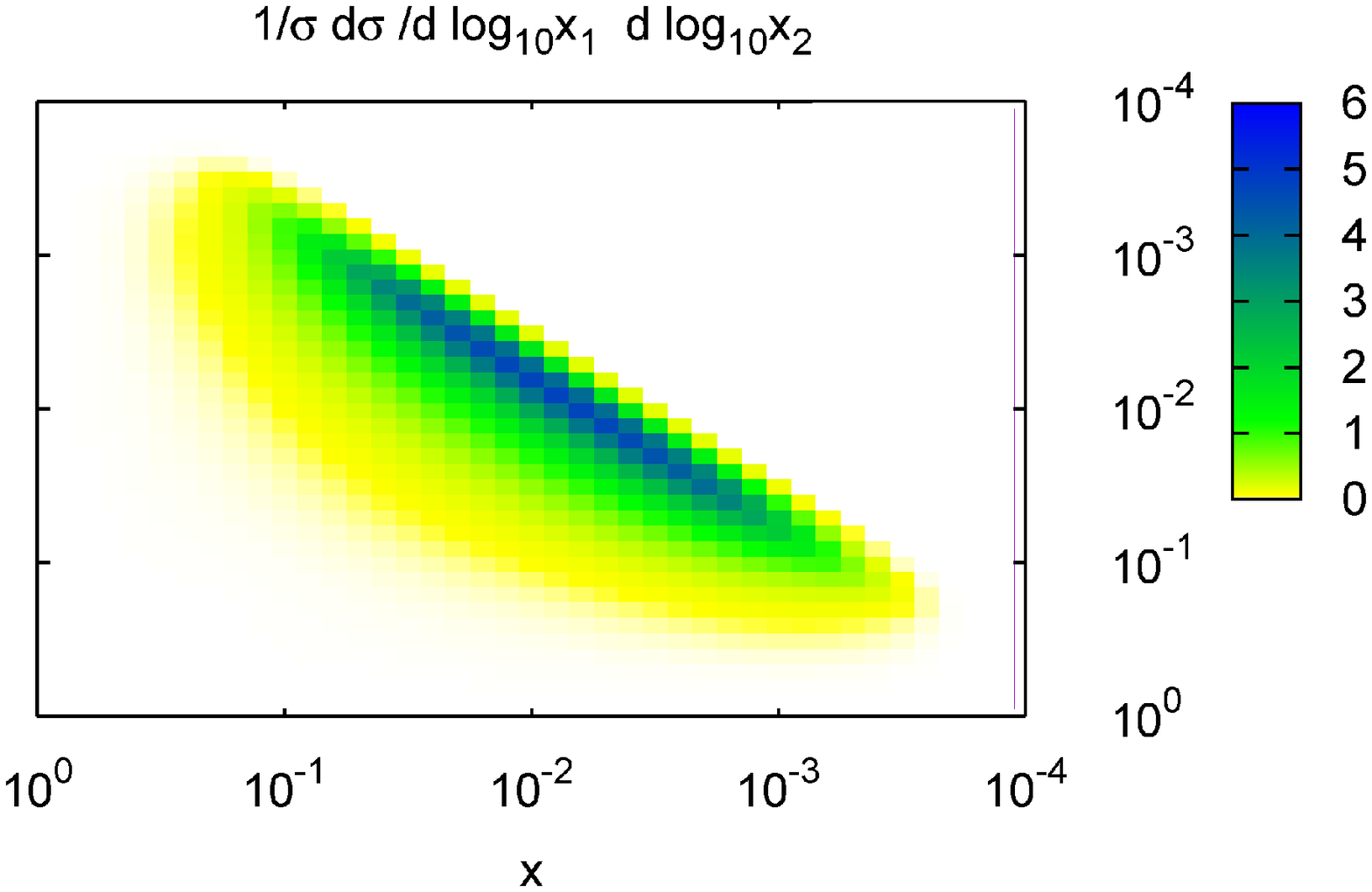}
\includegraphics[width=7.2cm]{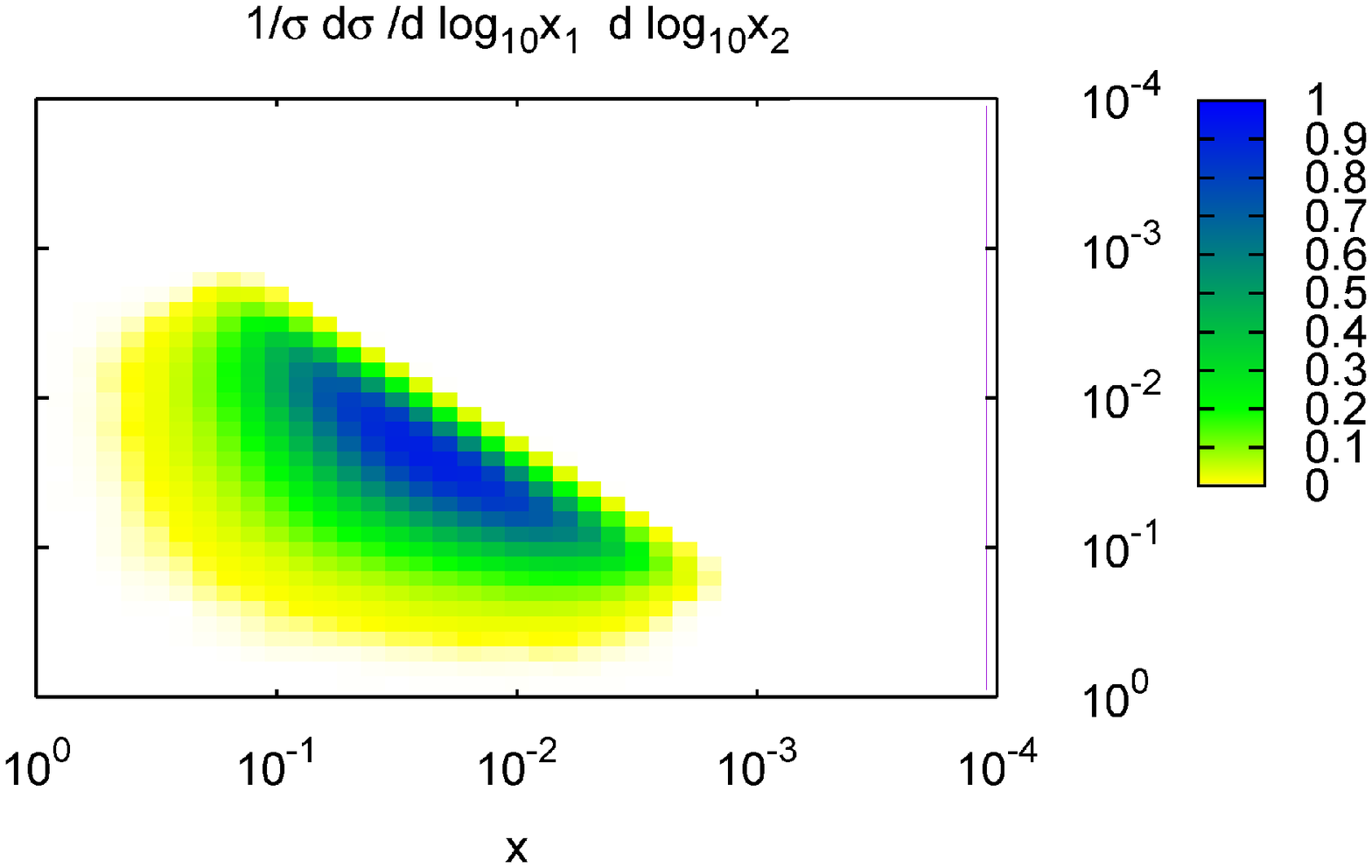}
\caption{Normalized 2-dimensional cross sections as functions of the initial partons longitudinal momenta fractions $x_1$ and $x_2$. Left panel: $W^-+c$ production case. Right panel: $Z+b$ production case.}
\label{Fig6}
\end{center}
\end{figure}

\begin{figure}
\begin{center}
\includegraphics[width=7.0cm]{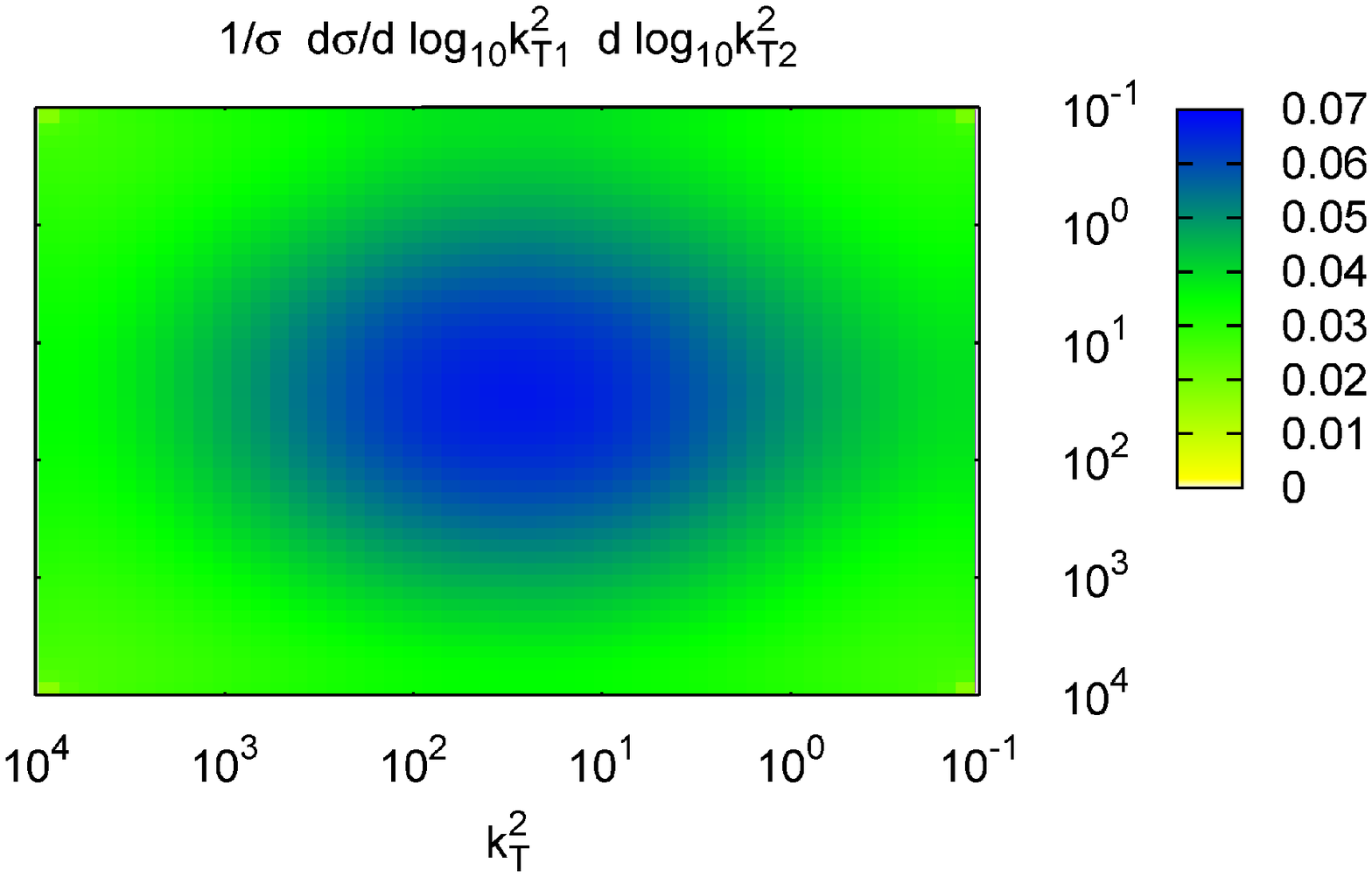}
\includegraphics[width=7.0cm]{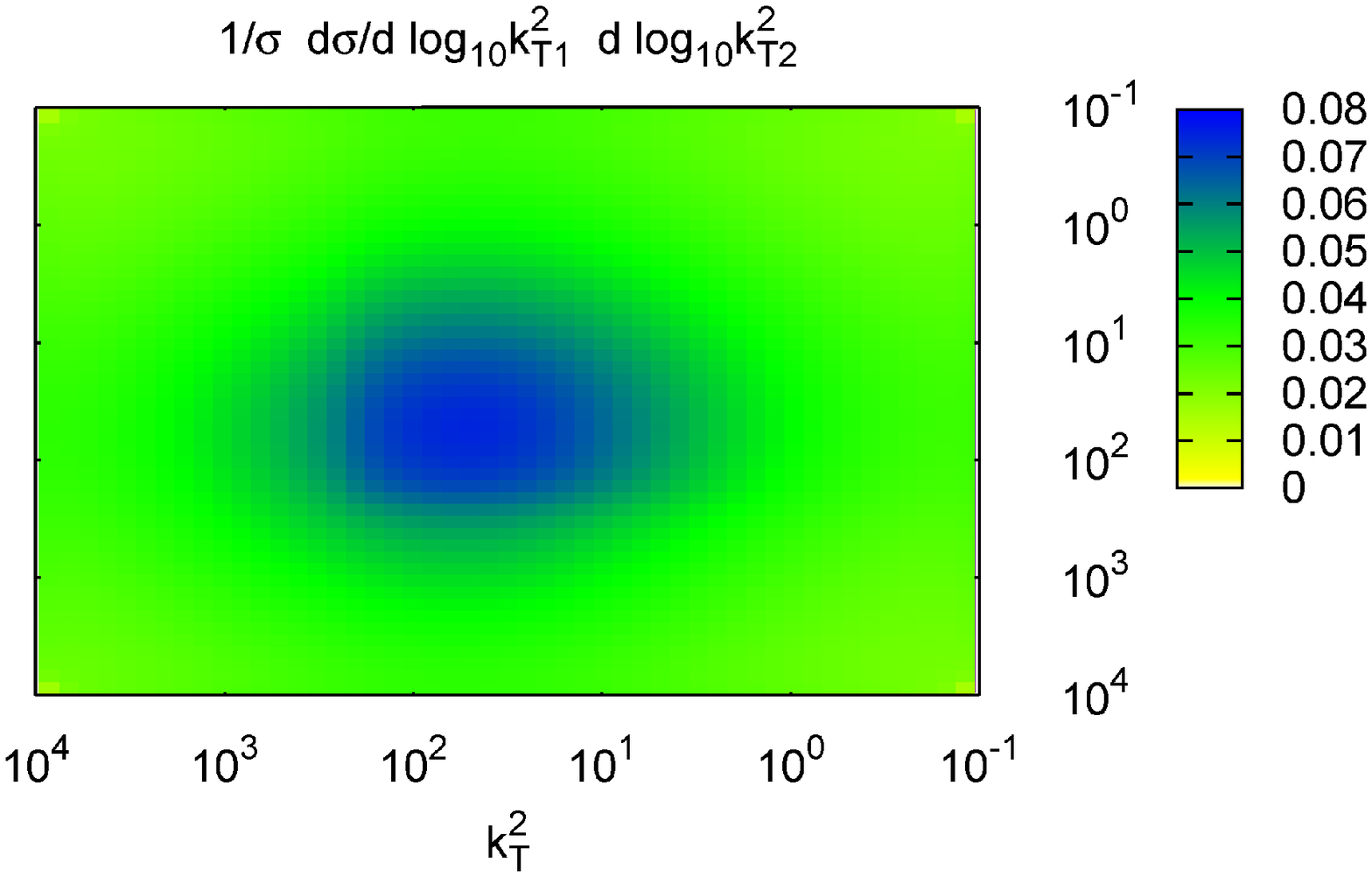}
\caption{Normalized 2-dimensional cross sections as functions of the initial partons transverse momenta squared $k_{T1}^2$ and $k_{T2}^2$. Left panel: $W^-+c$ production case. Right panel: $Z+b$ production case.}
\label{Fig7}
\end{center}
\end{figure}

\begin{figure}
\begin{center}
\includegraphics[width=7.0cm]{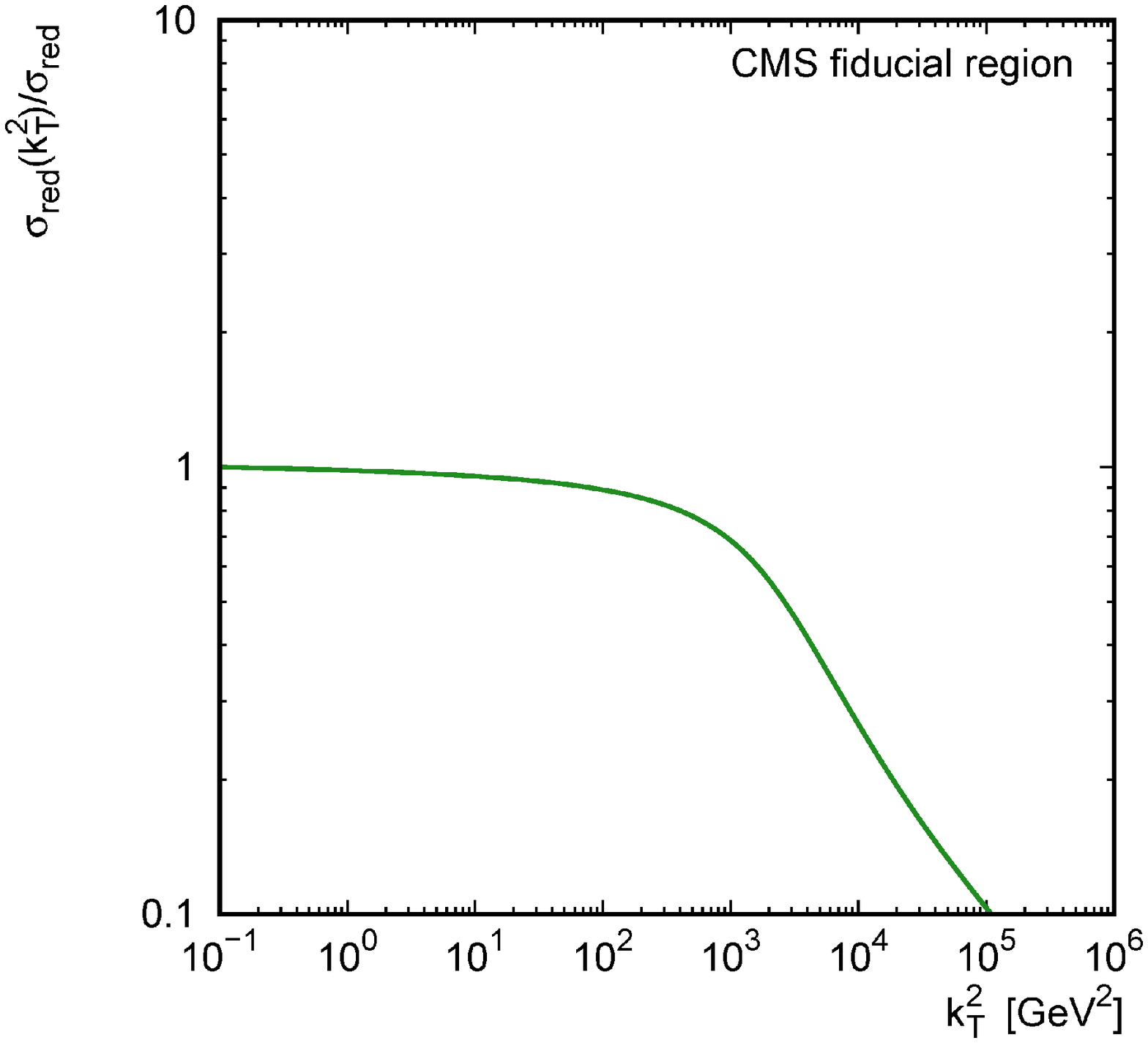}
\includegraphics[width=7.0cm]{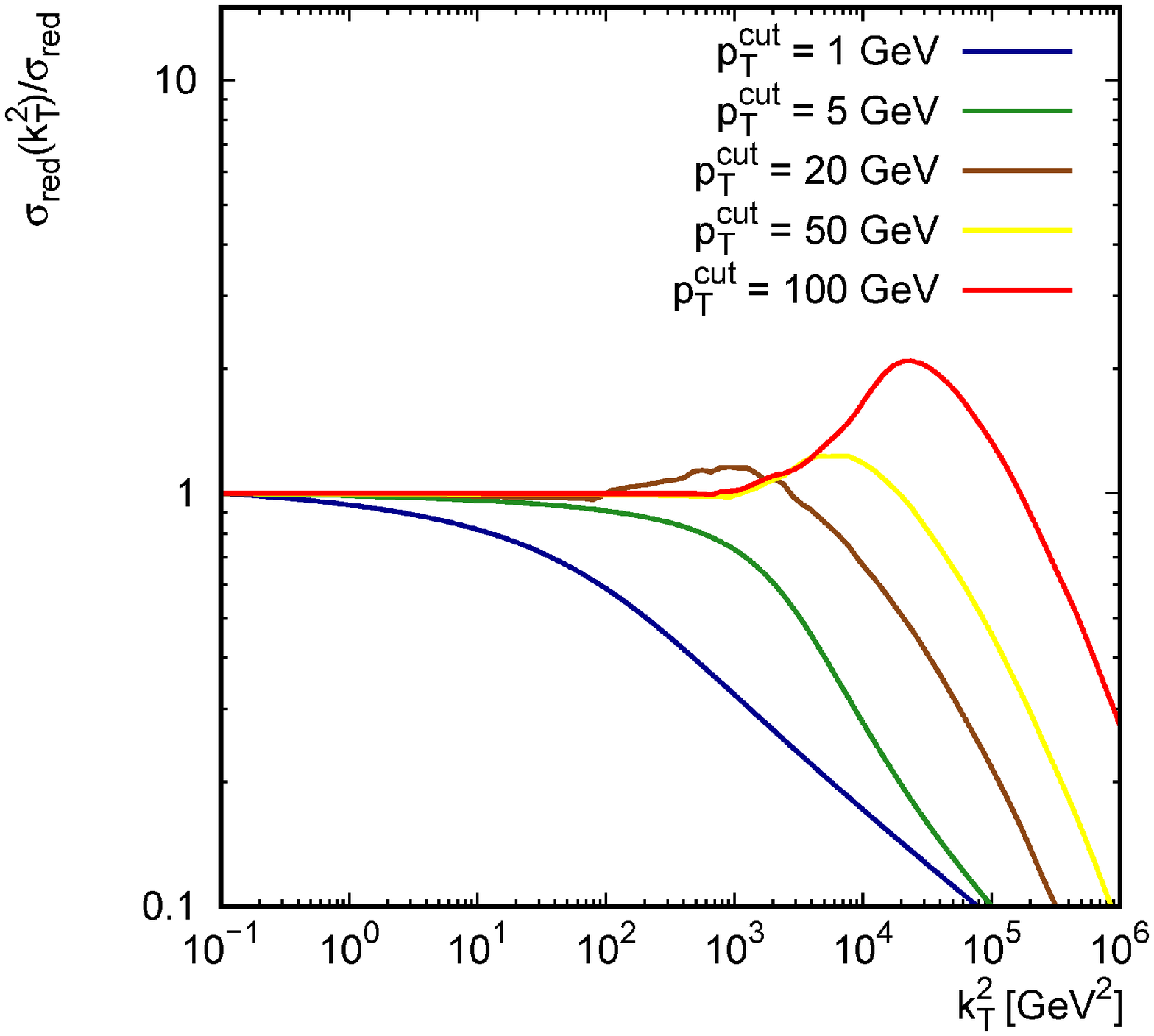}
\caption{The ratio of 'reduced' cross-sections, calculated with off-shell and on-shell matrix elements for the subprocess~(\ref{sg}) (see explainations in the text). Left panel: the ratio in experimental cuts of CMS. Right panel: only cuts on $p_T^c$ is kept in the range from 1 to 100~GeV.}
\label{Fig8}
\end{center}
\end{figure}

\begin{figure}
\begin{center}
\includegraphics[width=7.0cm]{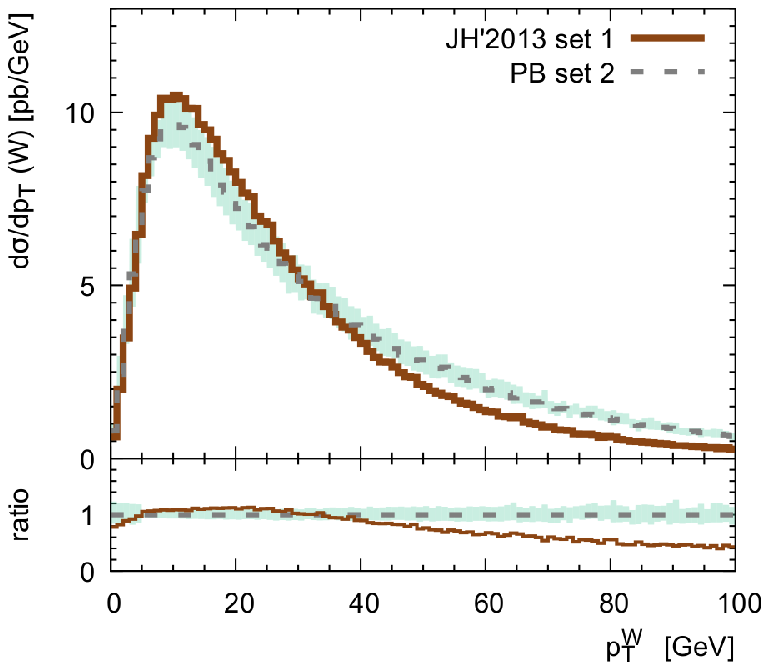}
\caption{Cross sections of $W^-+c$ production as functions of the $W$-boson transverse momentum. The notations are the same, as on the previous figure.}
\label{Fig5}
\end{center}
\end{figure}

\section*{Acknowledgements}
The authors thank S.P.~Baranov for useful discussions on the topic. A.V.L. and M.A.M. are grateful to DESY Directorate for the support in the framework of
Cooperation Agreement between MSU and DESY on phenomenology of the LHC processes
and TMD parton densities. M.A.M. was also supported by a grant of the foundation for the advancement of theoretical physics and mathematics "Basis" 17-14-455-1.

\end{document}